\documentclass{PoS}

\title{The Next Generation of IceCube Realtime Neutrino Alerts}

\ShortTitle{Realtime Neutrino Alerts}

\author{
The IceCube Collaboration\footnote{For collaboration list, see PoS(ICRC2019) 1177.}\\
{\itshape \href{http://icecube.wisc.edu/collaboration/authors/icrc19_icecube}{http://icecube.wisc.edu/collaboration/authors/icrc19\_icecube}}\\
E-mail: \email{ctung6@gatech.edu}
}

\abstract{

In 2016, IceCube initiated a system of public real-time alerts that are typically issued within one minute, following the detection of a neutrino candidate event that is likely to be of astrophysical origin.  The goal of these alerts is to enable multi-messenger observations that may identify the neutrino source. Through January 31, 2019, a total of 20 public alerts have been issued, with many of them receiving follow-up observations across multiple wavelength bands. One alert in particular, IceCube-170922A, was found to be associated with a flaring gamma-ray blazar, TXS 0506+056. This was the first >3 sigma association of a high-energy neutrino with an electromagnetic counterpart. In 2019, the IceCube collaboration is introducing a new set of neutrino candidate selections that expand the alert program. These new selections provide two alert channels. A "Gold" channel will issue alerts for neutrino candidates at least 50\% likely to be of astrophysical origin and is expected to deliver $\sim$10 alerts per year. Additionally a more frequent "Bronze" channel will provide $\sim$20 alerts per year for neutrino candidates that are between 30\% and 50\% likely to be of astrophysical origin. We present the neutrino event selections used to generate these alerts, the expected alert rates, and a description of the alert message.

\vspace{4mm}
{\bfseries Corresponding authors:}
Erik Blaufuss$^{1}$, Thomas Kintscher$^{2}$, Lu Lu$^{3}$, \speaker{Chun Fai Tung}$^{4}$\\
{$^{1}$ \itshape University of Maryland}\\
{$^{2}$ \itshape DESY, Zeuthen}\\
{$^{3}$ \itshape Chiba University}\\
{$^{4}$ \itshape Georgia Institute of Technology}

}

\FullConference{36th International Cosmic Ray Conference -ICRC2019-\\
		July 24th - August 1st, 2019\\
		Madison, WI, U.S.A.}

\begin{document}

\section{Introduction}\label{sec:intro}

High-energy neutrinos, being the "smoking gun" of high-energy hadronic interactions, play an important role in understanding the origin of cosmic rays. The discovery of the high-energy astrophysical neutrino flux by the IceCube Neutrino Observatory~\cite{Aartsen:2013jdh, Aartsen:2014gkd} marked the beginning of high-energy neutrino astronomy. A number of studies have been performed to locate point sources responsible for the astrophysical neutrino flux~\cite{Aartsen:2018ywr, Carver:2019icrc_pss}, but none has been found with 5$\sigma$ confidence level to date. 

IceCube is a cubic-kilometer neutrino detector~\cite{Aartsen:2016nxy} located at the South Pole. It does not observe neutrinos directly, but instead it has 5160 digital optical modules (DOMs) buried in the ice of Antarctica to capture the Cherenkov light emitted by the secondary particles produced in neutrino-nucleon interactions. The direction and energy of the neutrino can then be determined by reconstructing the secondary particles from the data collected by the DOMs. Therefore, the accuracy of the information of the neutrino depends greatly on the secondary particles.

The majority of the neutrino signals detected by IceCube can be classified into two categories: track events and cascade events. Track events are characterized by their long paths of light which can span multiple kilometers. They are caused by highly penetrating muons, which are produced as secondary particles in the charged current interactions of muon-flavour neutrinos (CC $\nu_{\mu}$). The long light paths allow more accurate reconstruction of the directions of the muons, and the uncertainty can be below 1 degree. However, because most of the track events lie only partly inside the instrumented volume, the energy reconstruction has large uncertainty. Cascade events are characterized by spheres of light. They can be caused by all kinds of neutrino interactions except CC $\nu_{\mu}$, when the interaction vertices are inside the instrumented volume. The reconstruction of cascade events is usually better in energy but worse in direction~\cite{Aartsen:2013vja}. The angular uncertainties are usually 10 to 15 degrees, but can also be much larger.

Multi-messenger astronomy combines the information provided by photons, neutrinos, and gravitational waves together to reveal the hidden physics in the most exotic environments of the universe. A potential neutrino source can be located if a counterpart in another messenger is observed in the direction from which an astrophysical neutrino candidate was detected. To achieve such observations, IceCube established a low-latency realtime neutrino alert system in April 2016~\cite{Aartsen:2016lmt}. When IceCube detects an astrophysical neutrino candidate which satisfies the selection criteria, a computer-readable message is generated and sent out to the community automatically. During its three years of operation, the average alert rate was $\sim$9 per year and the signal purity of the alert events was $\sim$25\%.

One specific IceCube alert, IC170922A, was sent out after a track like neutrino event was detected. Electromagnetic (EM) flares of different wavelengths were detected from the blazar TXS 0506+056 in spatial and temporal coincidence with the neutrino event by different telescopes, and the neutrino and gamma-ray observations were found to be correlated with 3$\sigma$ significance~\cite{IceCube:2018dnn}. IceCube also searched the archival data at TXS 0506+056's location and found a neutrino flare in the 2014-2015 season. This result shows that TXS0506+056 is a time-dependent neutrino source with $3.5\sigma$ significance~\cite{IceCube:2018cha}.

The success of the follow-up observations of TXS 0506+056 demonstrated the potential of the multi-messenger approach to neutrino astronomy. After three years of operation, the realtime neutrino alert system has shown room for improvement, which includes: (1) provide more neutrino candidates from a larger sample pool, (2) avoid mis-characterised events, (3) improve alert messages' clarity, and (4) define "signalness" for all the alerts.

To address these issues, updates have been performed on most parts of the realtime alert system. These updates include an expanded and improved event selection, which is discussed in Section~\ref{sec:selection}. In addition, the alert message format and alert streams are revamped to improve clarity and reduce confusion for the general astronomy community, which is discussed in Section~\ref{sec:message}. These updates result in a higher rate of alerts along with a higher signal purity, which is tabulated in Section~\ref{sec:rate}.


\section{IceCube Realtime Alert System Update}\label{sec:System}

The infrastructure of the realtime alert system remains largely the same as the previous system, which is described in detail in~\cite{Aartsen:2016lmt}. 
\begin{figure}
\begin{center}
\resizebox{0.8\textwidth}{!}{\includegraphics{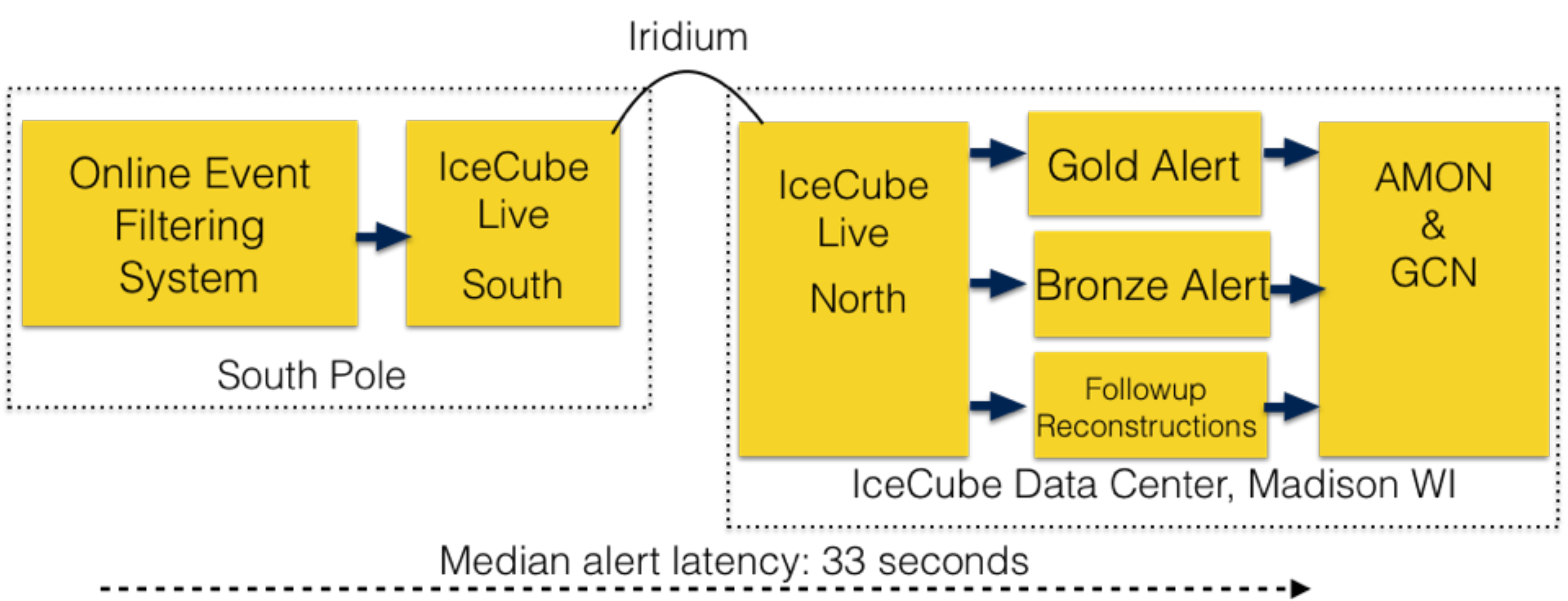}}
\end{center}
\caption{Schematic overview of the realtime alert system. At South Pole, information of the events satisfying the selection criteria is sent North instantly through the Iridium satellite system. In the North, the signalness of each event is assessed, and is used to decide if an alert is sent out. If the signalness is above 50\%, it is sent out via the Gold stream. If it is below 50\% but above 30\%, it is sent out via the Bronze stream. Both streams distribute the information in GCN Notice format.}
\label{fig:overview}
\end{figure}
As illustrated in Figure~\ref{fig:overview}, when IceCube detects an event, it is first processed through the filtering system. For events that pass the filter, they are sent to the IceCube data center over the Iridum satellite. After reaching the North, the remaining selection criteria are applied on the event to determine if it is an astrophysical neutrino candidate. If it is selected as a candidate, it is sent as an alert through either one of the two streams, namely "Gold" or "Bronze". The choice of stream depends on the signalness of the event, which measures the probability of the event being caused by an astrophysical neutrino. Signalness is defined as:
\begin{equation}
    Signalness(E,\delta) = \frac{N_{signal}(E,\delta)}{N_{signal}(E,\delta)+N_{background}(E,\delta)},
    \label{eq:signalness}
\end{equation}
where $N_{signal}$ and $N_{background}$ are the number of signal events and number of background events at declination $\delta$ above the selection-specific energy proxy $E$ . For example, $E$ can be the estimated neutrino energy. Candidates with signalness above 30\% but below 50\% are sent out in the Bronze stream, while candidates with signalness above 50\% are sent out in the Gold stream. Alerts from both streams are distributed as Gamma-ray Coordinates Network (GCN) Notices. A more sophisticated reconstruction for the neutrino candidate is performed in the North to provide more accurate direction and energy. This information is sent out in GCN circulars generally a few hours after the initial GCN notice. 

\subsection{Event Selection}\label{sec:selection}
The updated event selection scheme consists of three different selections, which are named GFU, HESE, and EHE. As a whole, they are responsible for picking out the neutrino-like track events most likely to be astrophysical from all of IceCube's data. HESE and EHE selections were already implemented in the previous realtime alert system. In the new realtime alert system, the HESE selection is updated while the EHE selection is kept unchanged. The GFU selection is a new addition to the lineup of selection criteria, and by employing a different strategy, it is able to pick up neutrino events which are missed by the other two selections. All of the selections are based on previous IceCube analyses and optimized for fast execution as filters in the online alert system.

\paragraph{GFU Track Selection}
The Gamma-ray Follow Up selection is designed to select high-quality track events that are similar to the ones used in IceCube's point-source analysis~\cite{Carver:2019icrc_pss, Aartsen:2016qbu}. Boosted decision trees (BDTs) are trained to pick out through-going track events that are caused by astrophysical neutrinos and are well reconstructed. To further improve the astrophysical purity, alerts are sent out only for the events with the highest energy. Energies of events from the Northern sky (up-going) are gauged with their reconstructed muon energies, and two thresholds are applied on them to obtain the 30\% and 50\% signalness. Energies of events from the Southern sky (down-going) are estimated with the total charge of the photoelectrons (PE) recorded by the detector. Since the background rate in this portion of the sky is strongly declination-dependent, a two-dimensional selection criteria which depends on both the declination and charge is applied to achieve the same signalness cut as the up-going events. 

\paragraph{HESE Track Selection}
High-Energy-Starting-Event (HESE) employs a vetoing technique to select neutrino events with the interaction vertices lying inside the detector~\cite{Aartsen:2014gkd}. However, the majority of starting events are cascades, which are not ideal for source-pointing due to their large angular uncertainty. Therefore, for the alert selection, additional cuts are applied to choose starting events with an out-going muon track. First, the likelihood of the event reconstruction must be in favor of the track-like hypothesis. Second, the measured track length must be longer than 200 meters, which also ensures the reconstruction quality. Finally, a declination-based cut on the charge of the recorded PE is applied to improve the astrophysical purity and to cut on the 30\% and 50\% signalness for the Bronze and Gold streams.

\paragraph{EHE Track Selection}
The Extreme-High-Energy (EHE) track selection is modified from the event selection used in the analysis which led to the observation of the first PeV neutrinos~\cite{Aartsen:2013bka}. It targets track-like neutrino events with energy between 500 TeV and 10 PeV by requiring the passing events' numbers of photoelectron (NPE) to be at least 4000. A fit quality parameter ($\chi^2$) is used to select well-reconstructed track events, and a declination-dependent cut on NPE is used to increase the portion of astrophysical events. The NPE cut value is set to achieve 50\% signalness in the final sample, which makes EHE track events exclusively in the Gold stream.

\begin{figure}
\begin{center}
\resizebox{0.7\textwidth}{!}{\includegraphics{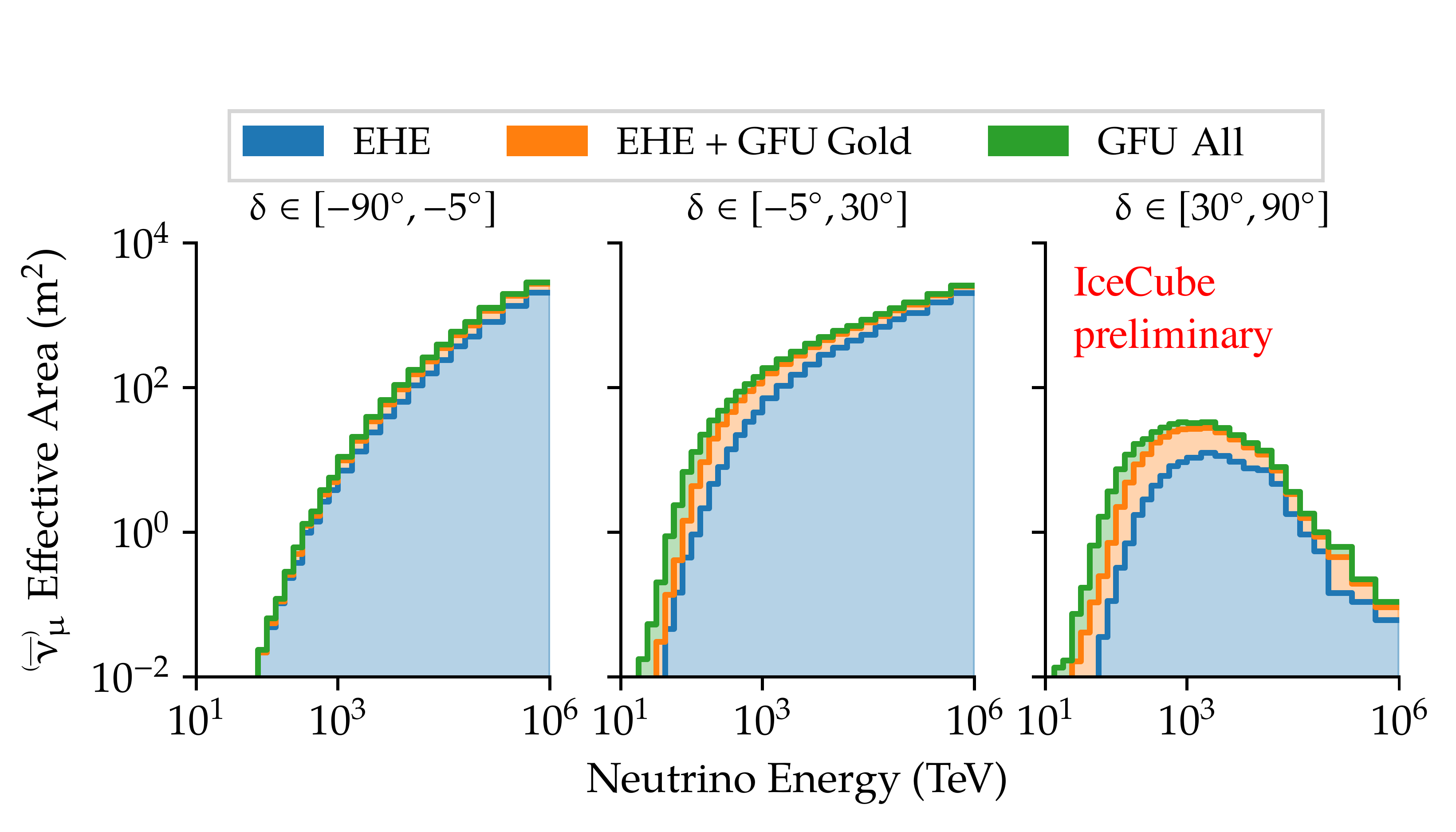}}
\resizebox{0.7\textwidth}{!}{\includegraphics{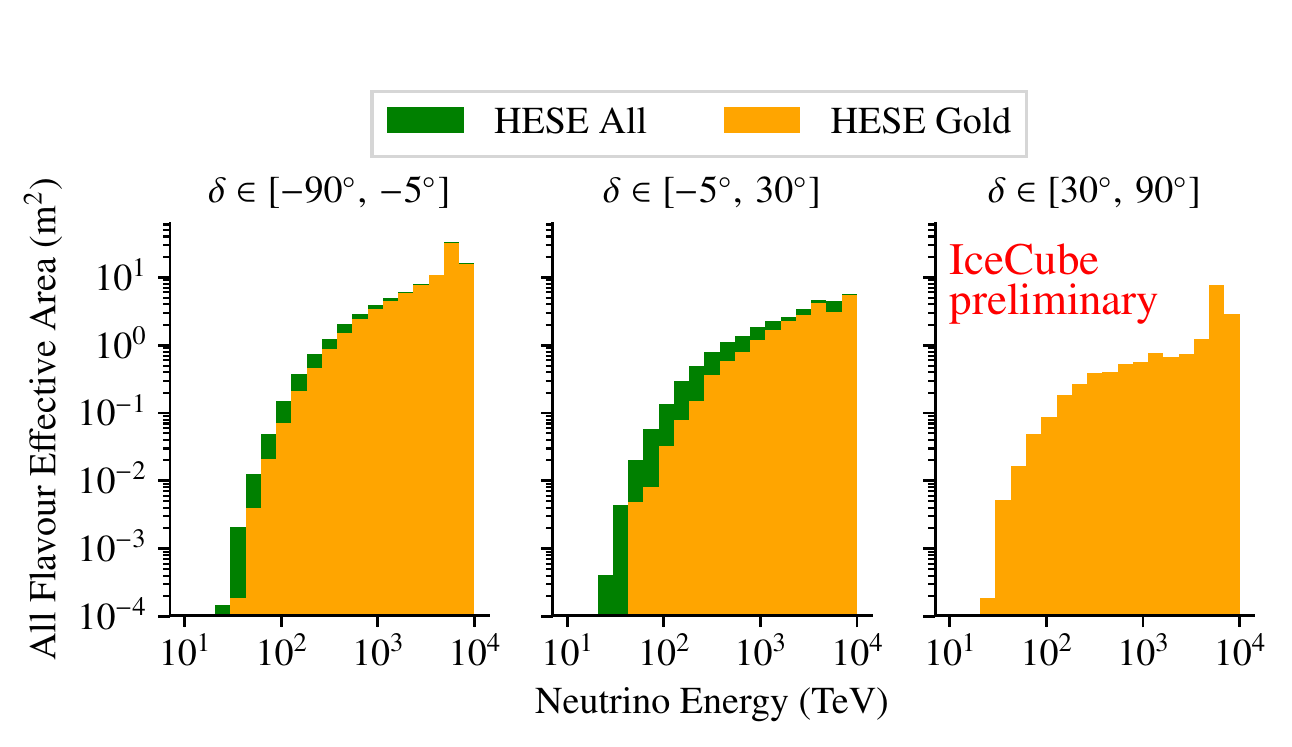}}
\end{center}
\caption{Effective areas of the new realtime neutrino alerts as a function of neutrino energy. Top: Muon neutrino effective area of selections that target through-going neutrino (EHE + GFU) in three declination bands. The effective area of the EHE selection is also shown for reference. Bottom: Neutrino effective area of starting track events selection (HESE) in three declination bands.}
\label{fig:effA}
\end{figure}

The effective area of these selections is shown in Figure~\ref{fig:effA}. It is possible that an event passes multiple selections, and each returns a different set of values for the event. The most probable overlap is between the GFU selection and the EHE selection, because they both target through-going track events. However, only one alert is sent out. The source of information reported follows a hierarchical order: GFU first, EHE second, and HESE last. This ordering is chosen based on the signal purity and angular resolutions of the three selections.

\subsection{Alert Message and GCN Circular}\label{sec:message}

\begin{figure}
\begin{center}
\resizebox{0.45\textwidth}{!}{\includegraphics{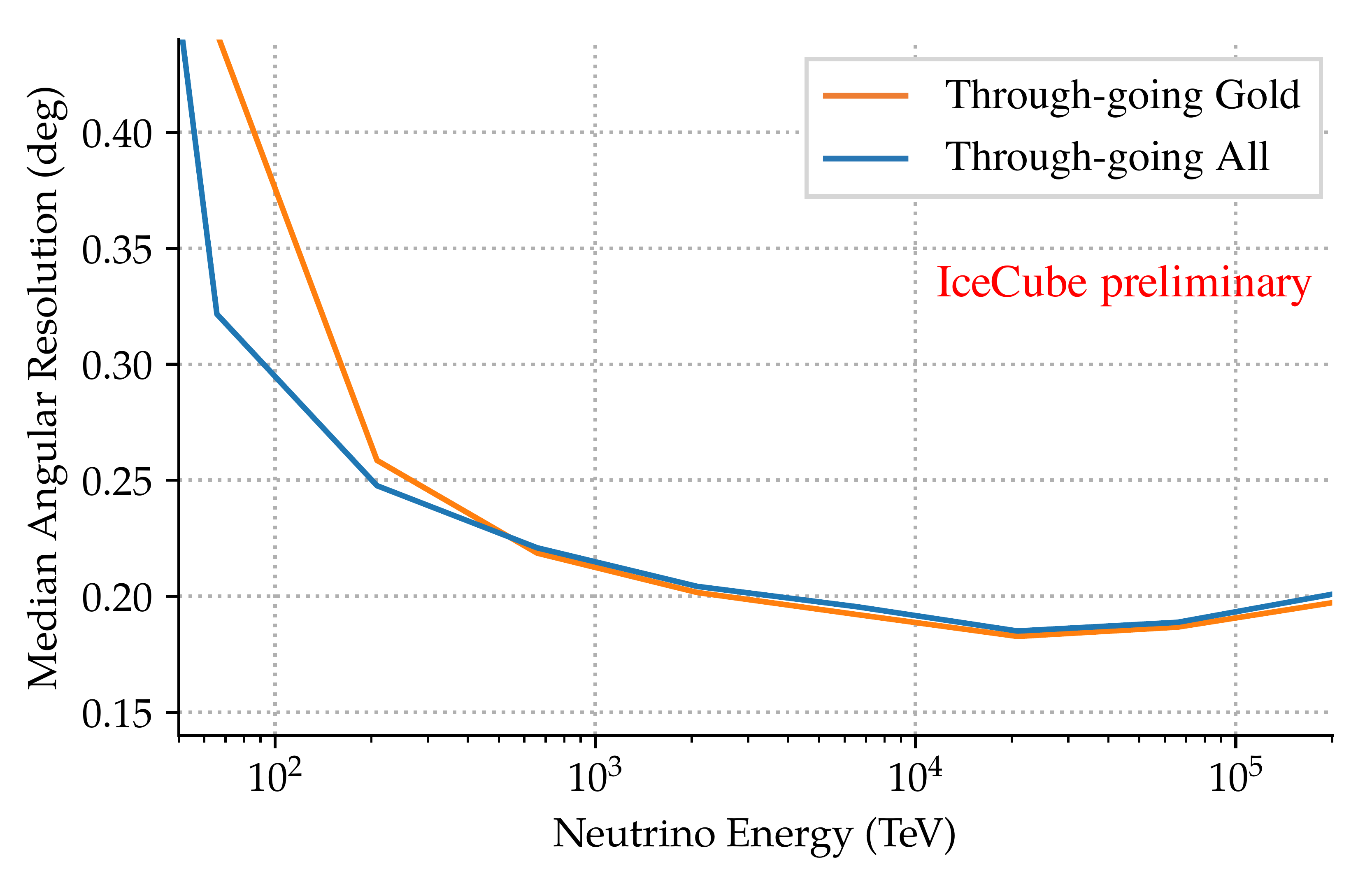}}
\resizebox{0.45\textwidth}{!}{\includegraphics{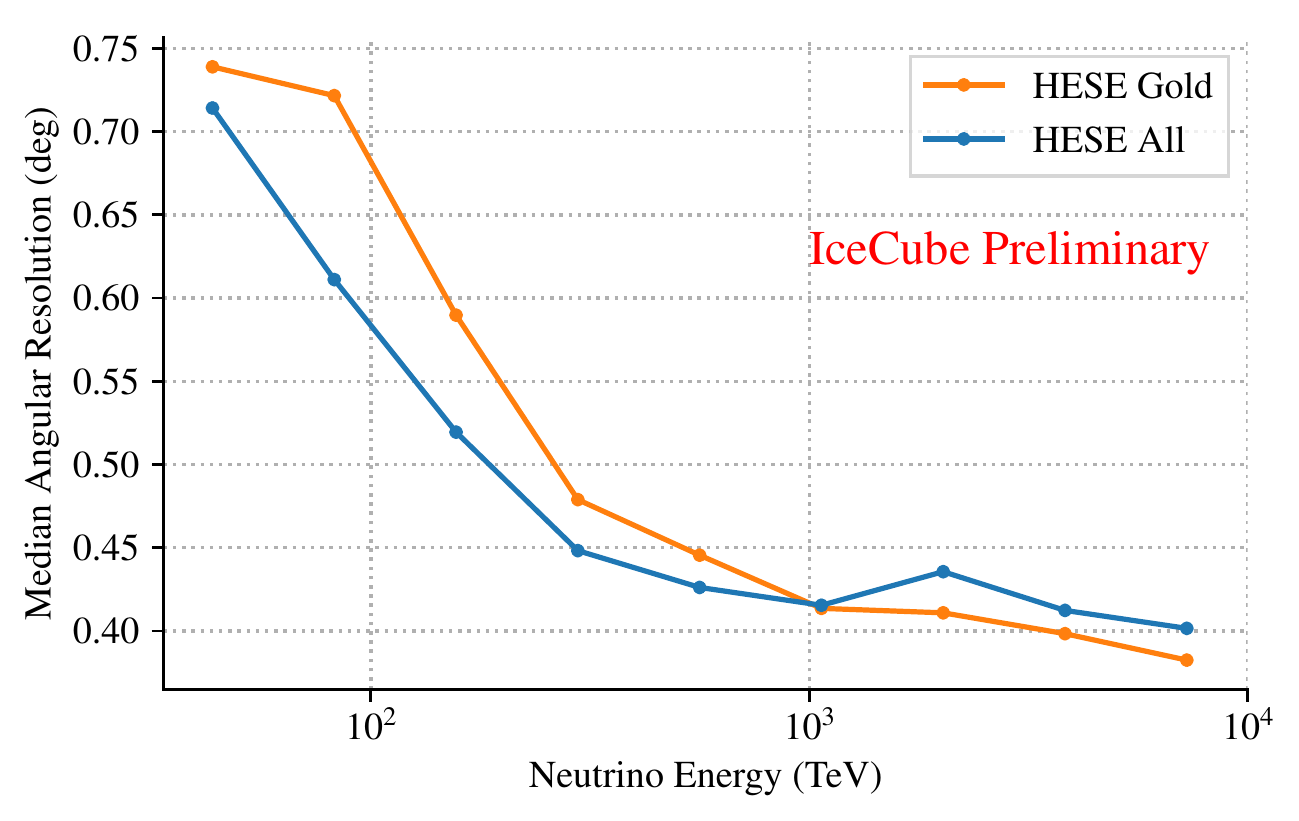}}
\end{center}
\caption{Angular resolution of IceCube realtime neutrino alerts as a function of neutrino energy. Left: Angular resolution of through going track events (GFU and EHE selections). Right: Angular resolution of starting track events (HESE selection). In this figure, "all" includes all the events above the Bronze parameter cut in the corresponding selection, and "Gold" includes all the events above the Gold parameter cut, so the Gold events are counted in both curves. The Gold alerts have a worse angular resolution at lower energy region. For through-going tracks, this is because the events have high declination, where the reconstruction is more difficult . For starting tracks, this is because the events have bright cascades at the interaction vertices. Also, as a new practice introduced in this update, the angular error reported in the GCN Notice is set to be at least 0.2 degrees to account for systematic uncertainty. }
\label{fig:angularerr}
\end{figure}

The content of the message is redesigned to improve the ease of understanding the details and significance of the neutrino candidate. Each GCN Notice will include the following:
\begin{itemize}
    \item \textbf{Discovery time and date} - 0.01 second precision expressed in UTC format.
    \item \textbf{IceCube run number and event number} - unique IDs used within the IceCube Collaboration.
    \item \textbf{Right Ascension and Declination} - the reconstructed direction from which the neutrino candidate came. Values in J2000, current, and 1950 epochs are reported.
    \item \textbf{Direction Error} - angular radii of the 50\% and 90\% containment circles. Errors are derived from Monte Carlo simulations of similar events. The 50\% error as a function of neutrino energy is shown in Figure~\ref{fig:angularerr}. 
    \item \textbf{Signalness} - probability of the neutrino candidate being an astrophysical neutrino. Definition follows Eq~\ref{eq:signalness}.
    \item \textbf{False Alarm Rate} - yearly rate of alerts that have equal or higher signalness.
    \item \textbf{Likely Neutrino Energy} - the most likely energy of the neutrino deduced from the parameters of the alert event under the astrophysical neutrino hypothesis. The spectrum of the diffuse astrophysical neutrino flux is assumed to be a power law with spectral index -2.19.
\end{itemize}
All the values for the quantities listed above (except time and date and run/event number) are calculated based on the event sample and historical observations of the selection passed by the neutrino candidate. All of the information is generated during the reconstructions performed at South Pole.

Since the alerts are designed to promote follow-up observation instead of detailed analysis, the potential systematics of the information listed above are not included in the message. For example, the reported signalness and neutrino energy vary with spectral index of the astrophysical neutrino flux. It must also be noted that because a significant portion of a muon track can lie outside of the detector's instrumented volume, the true neutrino energy can be much higher than the reported value.

The reconstruction methods applied at South Pole are limited by the computation power, so they employ assumptions to improve execution speed. When the IceCube data center receives the complete data of the events, a more sophisticated reconstruction method is applied to them to obtain a more refined direction and angular uncertainty. Once the reconstruction is finished and verified, the new information is sent out via GCN Circular. The typical time lag between the automated GCN Notice and the GCN Circular is a few hours.
    
\subsection{Expected Alert Rate}\label{sec:rate}
The expected rates of signal events passing the updated selection are calculated using the best-fit diffuse astrophysical neutrino flux, which has a spectral index of $-2.19$ and normalization at 100 TeV of $1.01\times10^{-18} \mathrm{GeV^{-1}cm^{-2}s^{-1}sr^{-1}}$, as reported in \cite{Haack:2017dxi}. The rate of background events passing is calculated with the simulated atmospheric contamination, which includes both neutrinos and muons. These expected values are then compared with the observed values obtained by applying the same event selection to seven years of IceCube data. The expected and observed rates of passing each selection are tabulated in Table~\ref{tab:rates}. 

\begin{table}[ht]
\begin{centering}
\begin{tabular}{|l|l|l|}
\hline
                         & Gold Events & Bronze Events \\ \hline
Signal ($E^{-2.19}$)     & 6.6 (Total) &  2.8 (Total) \\ 
                                          & 5.1 (GFU)    &  2.5 (GFU) \\
                                          & 0.5 (HESE)  & 0.3 (HESE) \\
                                          & 2.1 (EHE)    &  \\ \hline
Atmospheric Backgrounds  & 6.1 (Total) & 14.7 (Total) \\
                                          & 4.7 (GFU)    & 13.8 (GFU) \\
                                          & 0.4 (HESE)  & 0.9 (HESE) \\
                                          & 1.9 (EHE)    &   \\ \hline
Observed historical rate & 9.9 (Total) & 19.5 (Total) \\ 
                                          & 7.8 (GFU)    & 18.4 (GFU) \\
                                          & 1.1 (HESE)   & 0.9 (HESE) \\
                                          & 4.3 (EHE)    & \\ \hline
\end{tabular}
\caption{Expected and observed passing rates for Gold and Bronze selections. All values shown are events per year. Because of the overlap between GFU and EHE, the total rate of Gold alerts is not the sum of all selections.}
\label{tab:rates}
\end{centering}
\end{table}

\begin{figure}
\begin{center}
\resizebox{0.45\textwidth}{!}{\includegraphics{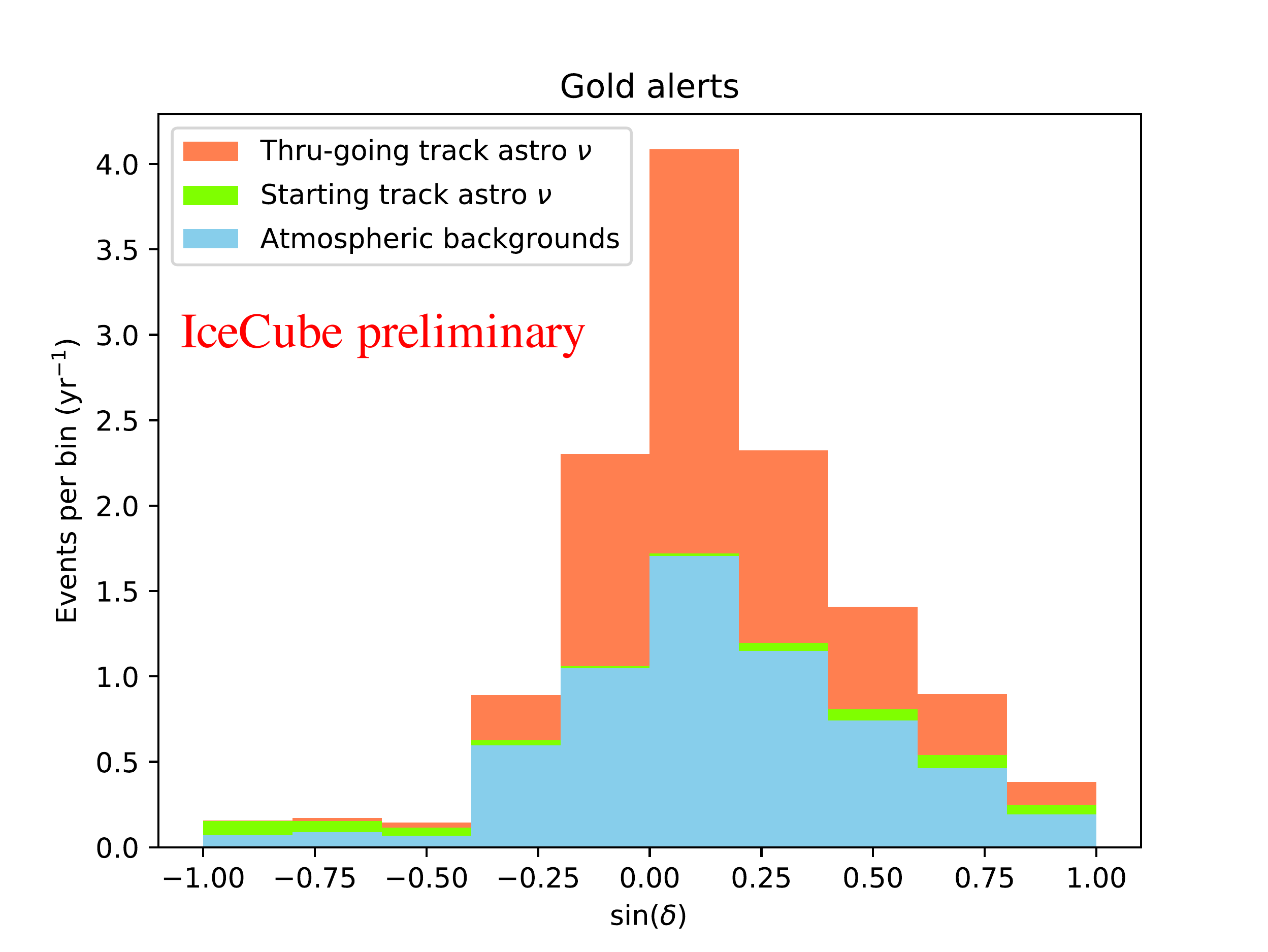}}
\resizebox{0.45\textwidth}{!}{\includegraphics{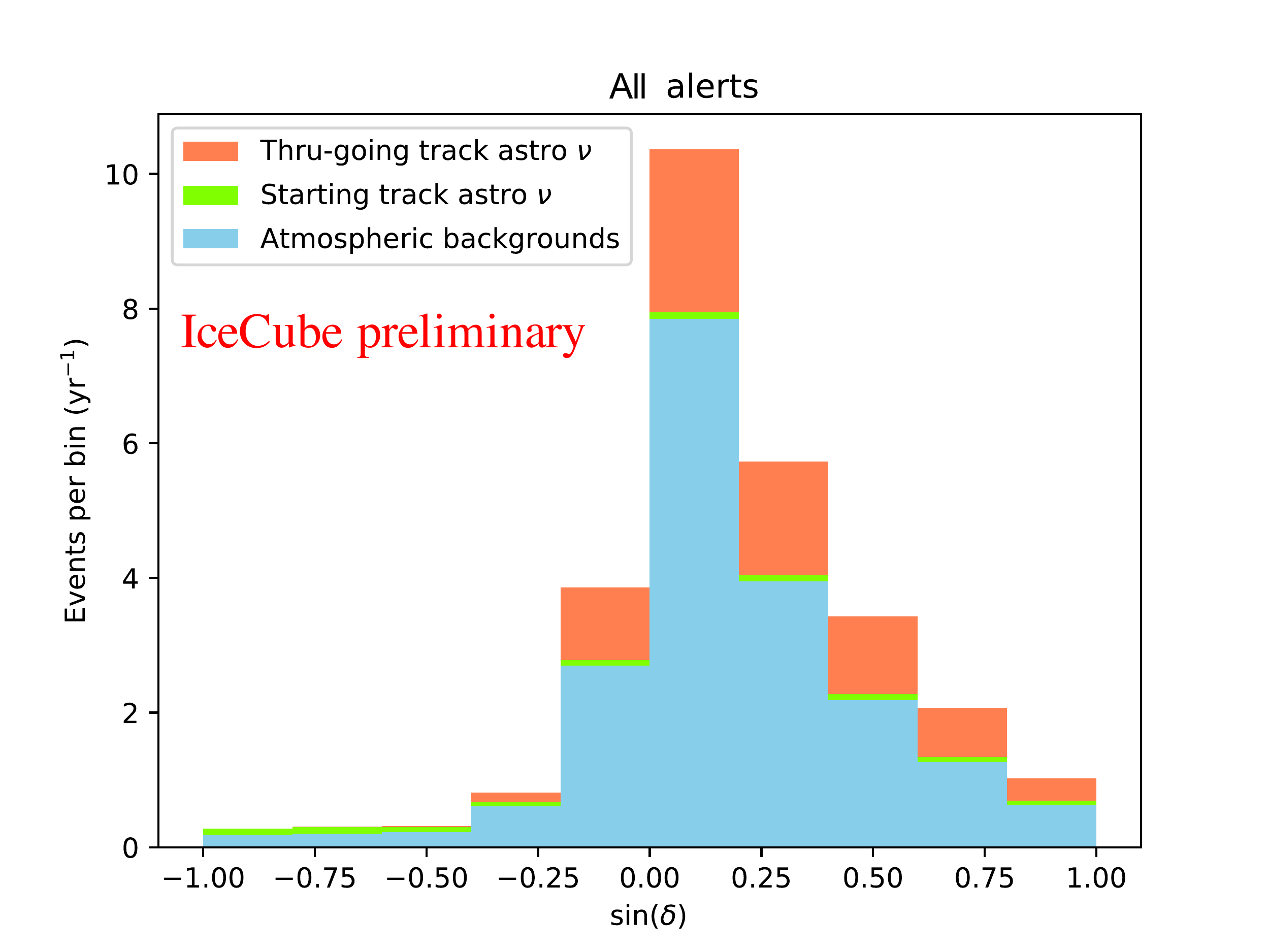}}
\end{center}
\caption{Expected yearly rate of IceCube realtime alerts distribution in declination. "Thru-going track astro $\nu$" includes events that passed either GFU or EHE selection; "starting track astro $\nu$" includes events that passed HESE selection. Left: Alerts passed the Gold criteria. Right: Alerts passed either the Gold or Bronze criteria.}
\label{fig:decdist}
\end{figure}

The majority of the alerts are expected to be through-going tracks (from GFU and EHE), with a small fraction contributed by starting tracks (from HESE). These alerts are not expected to be distributed isotropically in declination. This is a consequence of the declination dependence of background events and high-energy neutrinos' Earth absorption effect. As shown in Figure~\ref{fig:decdist}, most alerts are located within a 30-degree range centered north of the celestial equator.

\section{Summary and Future Outlook}
To help realize the potential of multi-messenger astronomy, IceCube has designed a new realtime neutrino alert system. This new system has been deployed since June 17th, 2019. After its deployment, IceCube is now issuing realtime alerts at an increased rate, and these alerts have a higher astrophysical purity and more reliable information than the previous generation. These changes provide more opportunities for follow-up observations to the astronomical community.

As multi-messenger astronomy becomes more mature, it is  reasonable to expect more frequent coincident observations in the future. The next multi-messenger detection might help resolve the mysteries of the origin of ultra-high energy cosmic rays, or it might give us another riddle about the high-energy universe.

\bibliographystyle{ICRC}
\bibliography{references}

%

\end{document}